\pdfoutput=1

\newcommand{\Gmat}[0]{\boldsymbol{G}}
\newcommand{\Xhat}[0]{\boldsymbol{\hat{X}}}

\newcommand{\Ygrid}[0]{\boldsymbol{Y}_{grid}}
\newcommand{\Tmat}[0]{\boldsymbol{\mathcal{T}}}

\newcommand{\metricA}[0]{$\text{ER}_{\text{LD}}\downarrow$}
\newcommand{\metricB}[0]{$\text{F}_{\text{LD}}\uparrow$}
\newcommand{\metricC}[0]{$\text{LE}_{\text{CD}}\downarrow$}
\newcommand{\metricD}[0]{$\text{LR}_{\text{CD}}\uparrow$}
\newcommand{\metricE}[0]{$\rm{\mathcal{E}_{SELD}}\downarrow$}

\documentclass{article}

\usepackage{amsmath,amssymb, graphicx}
\usepackage{spconf} 
\usepackage{xcolor}
\usepackage[colorlinks=true, allcolors=blue]{hyperref}
\usepackage{booktabs}
\usepackage{tabularx}
\usepackage{caption}
\usepackage{multirow}
\usepackage{mathtools}
\usepackage{marginnote}

\usepackage{eso-pic,xcolor}
\AddToShipoutPictureBG*{%
  \put(\LenToUnit{.1\paperwidth},\LenToUnit{4em})%
    {\parbox[3pt]{55em}{%
      \footnotesize\copyright 2021 IEEE. Personal use of this material is permitted. Permission from IEEE must be obtained for all other uses, in any current or future media, including reprinting/republishing this material for advertising or promotional purposes, creating new collective works, for resale or redistribution to servers or lists, or reuse of any copyrighted component of this work in other works.
      }%
    }%
}

\title{Spatial Mixup: Directional loudness modification as data augmentation for sound event localization and detection}
\name{Ricardo Falcón-Pérez$^{\star *}$, Kazuki Shimada$^{\dagger}$, Yuichiro Koyama$^{\dagger}$,  Shusuke Takahashi$^{\dagger}$,  Yuki Mitsufuji$^{\dagger}$}
  
  \address{$^{\star}$Aalto University, Espoo, Finland \\
      $^{\dagger}$Sony Group Corporation, Tokyo, Japan}
      
\begin{document}
\ninept
\maketitle
\begin{abstract}
Data augmentation methods have shown great importance in diverse supervised learning problems where labeled data is scarce or costly to obtain. For sound event localization and detection (SELD) tasks several augmentation methods have been proposed, with most borrowing ideas from other domains such as images, speech, or monophonic audio. However, only a few exploit the spatial properties of a full 3D audio scene. We propose Spatial Mixup, as an application of parametric spatial audio effects for data augmentation, which modifies the directional properties of a multi-channel spatial audio signal encoded in the ambisonics domain. Similarly to beamforming, these modifications enhance or suppress signals arriving from certain directions, although the effect is less pronounced. Therefore enabling deep learning models to achieve invariance to small spatial perturbations. The method is evaluated with experiments in the DCASE 2021 Task 3 dataset, where spatial mixup increases performance over a non-augmented baseline, and compares to other well known augmentation methods. Furthermore, combining spatial mixup with other methods greatly improves performance.
\end{abstract}

\renewcommand{\thefootnote}{\fnsymbol{footnote}}
\footnote[0]{
$*$Work done during an internship at Sony Group Corporation.}

\section{Introduction}
\label{sec:intro}

\begin{figure}[h!tb]
    \begin{minipage}[b]{0.98\linewidth}
      \centering
      \includegraphics[trim={0 2.5cm 0 2.5cm},clip, width=0.8\textwidth]{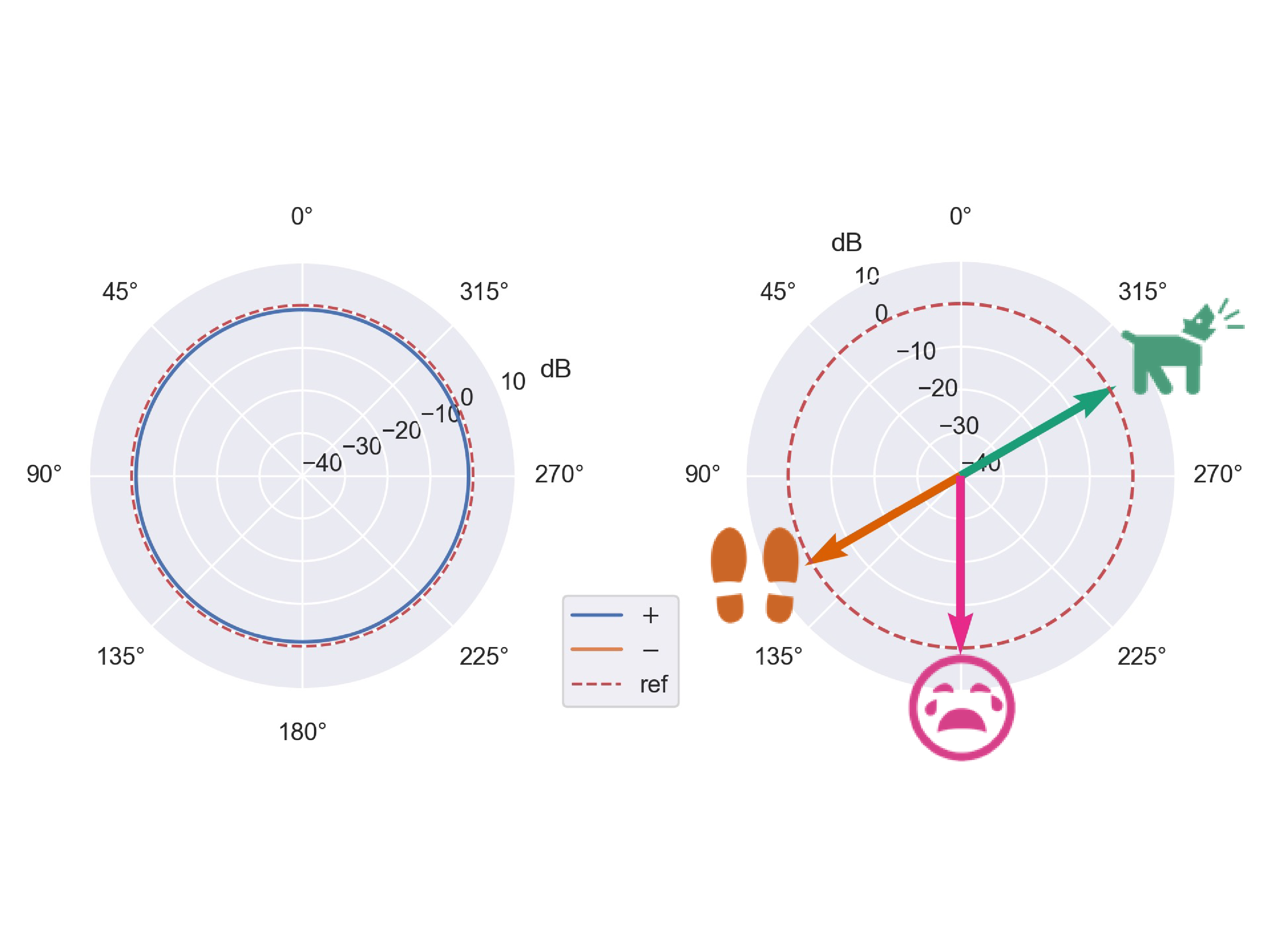}
    \end{minipage}
    \begin{minipage}[b]{0.98\linewidth}
      \centering
      \includegraphics[trim={0 2.5cm 0 2.5cm},clip, width=0.8\textwidth]{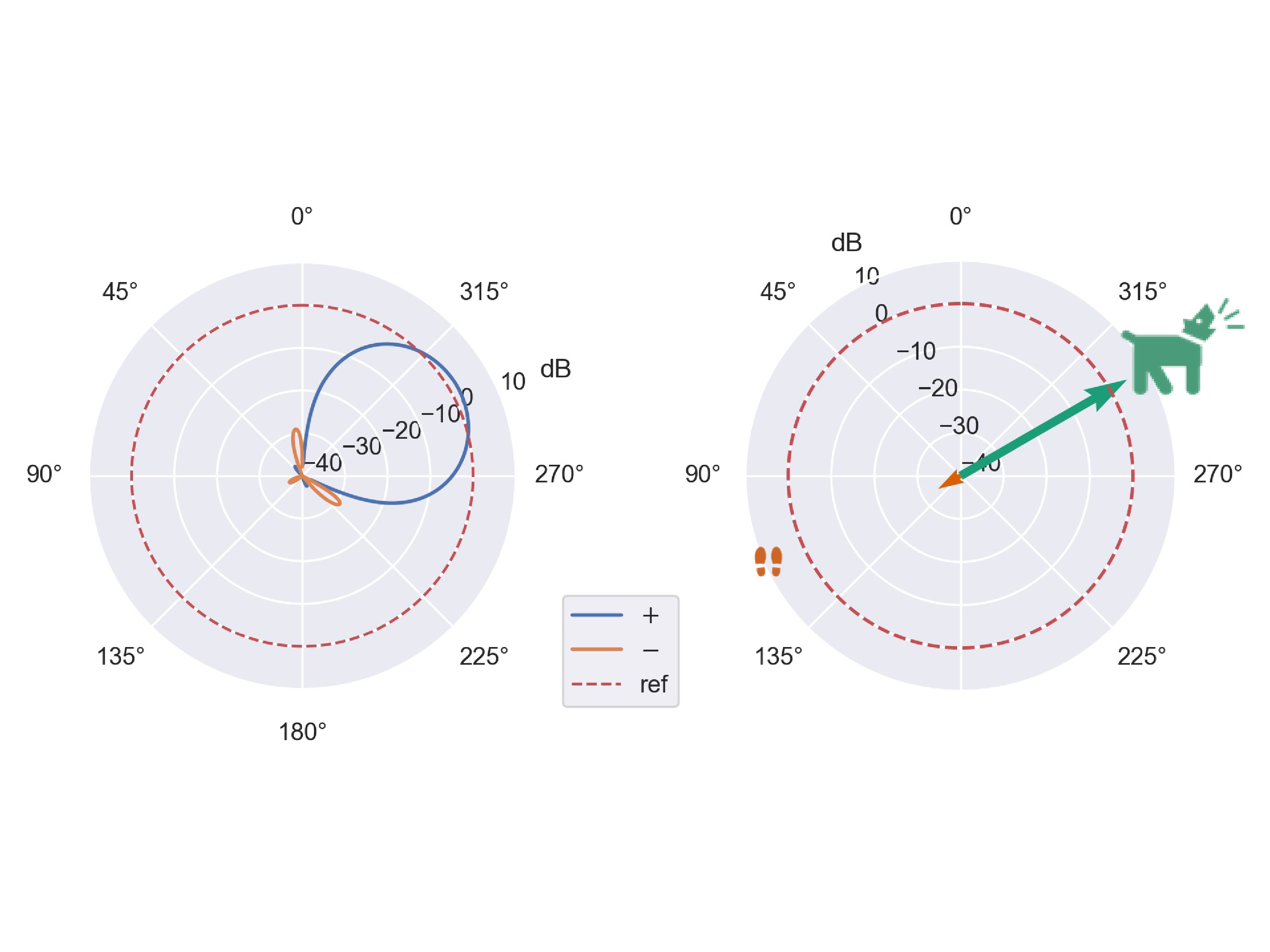}
    \end{minipage}
    \begin{minipage}[b]{0.98\linewidth}
      \centering
      \includegraphics[trim={0 2.5cm 0 2.5cm},clip, width=0.8\textwidth]{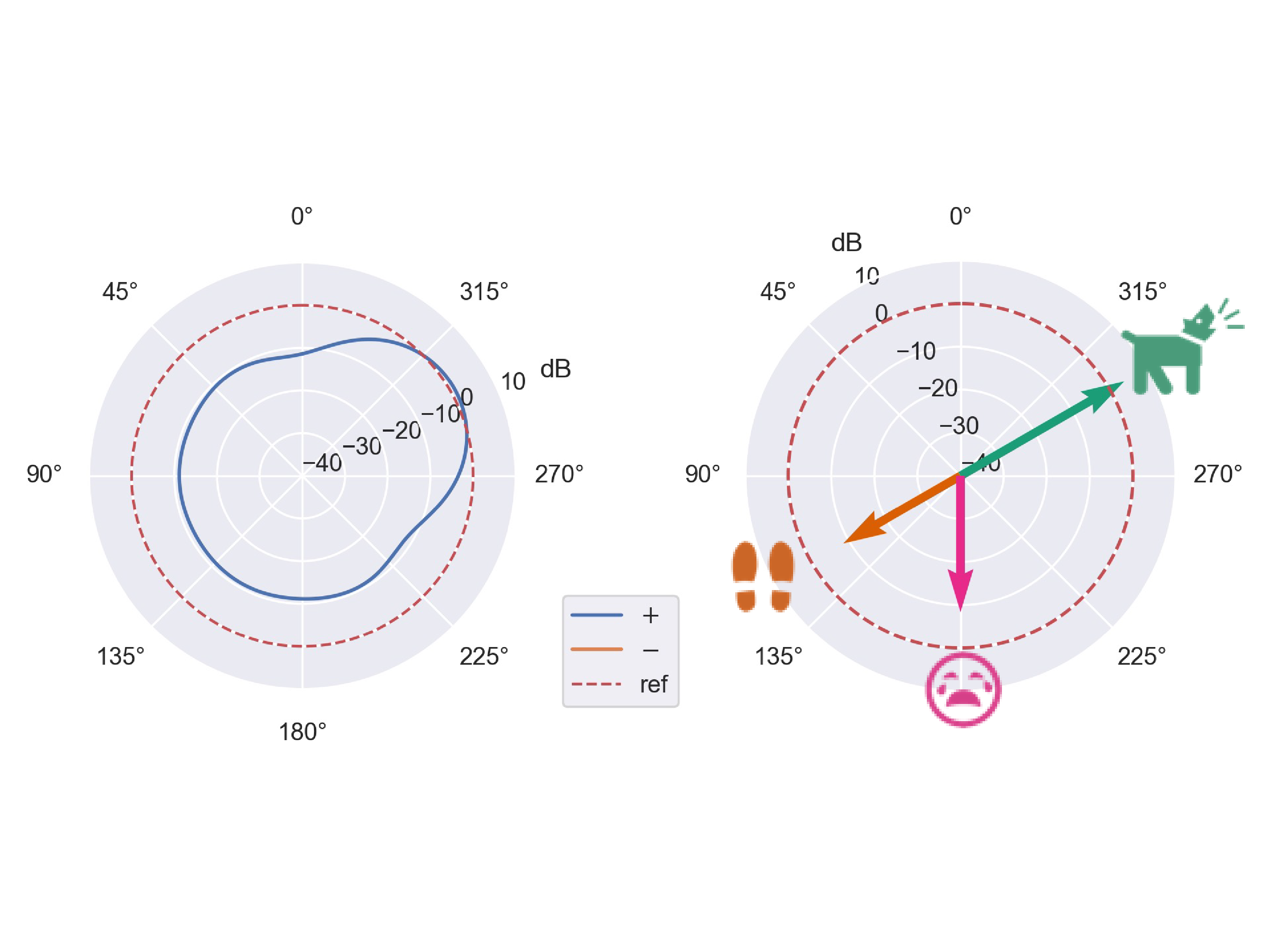}
    \end{minipage}
    \captionsetup{width=0.98\linewidth}
    \caption{Illustrative comparison of different spatial transformations for a 2D scene. (Top) shows the original omnidirectional response and (top right) the corresponding ACCDOA labels pointing to sources. (Middle left) shows a traditional beamformer and (middle right) the corresponding transformed ACCDOA labels, where high suppression effectively eliminates events outside the main lobe. (Bottom left) shows the proposed spatial mixup with a soft spherical cap and (bottom right) the optionally transformed labels.}
    \label{fig:approach}
\end{figure}

A sound event localization and detection (SELD) task is a dual task where the goal is to classify the type of sounds present in an acoustic scene as well as estimate their direction of arrival (DOA) \cite{Purwins2019, adavanne_2018:SoundEventLocalization, Adavanne2018DirectionOA}. It is similar to image classification and image segmentation in the sense that the objective is to identify the content and location of elements in the scene. So SELD can be thought of as a machine listening problem that has attracted significant attention lately, as it is key for artificial intelligence methods to understand the world via sound \cite{chen2020soundspaces}.

In the current state of the art, most SELD tasks are solved using complex systems that are based on deep learning networks \cite{Shimada_SONY_task3_report, Nguyen_NTU_task3_report, Lee_SGU_task3_report}. The input features are usually either log mel spectrograms (or other time-frequency transforms), or raw waveforms extracted from a soundfield recorded by a microphone array. The models vary in size and complexity, but include multiple architectures of convolutional networks with additional elements such as recurrent layers \cite{cakir2017convolutional}, transformer-based self-attention modules \cite{miyazaki2020weakly} or dense residual blocks \cite{takahashi2021densely}. Post-processing techniques such as weight averaging, ensemble methods and filtering are also common \cite{Shimada_SONY_task3_report}.

However, the labeled datasets available to train and evaluate such systems are generally very limited, as this requires labeling for both event class and DOA for each frame in the recording. Therefore, a significant component of modern SELD systems is the data augmentation strategy \cite{wang_2021:FourStageDataAugmentation}. Although spectrograms can be interpreted as a 2D image representation of sound, and systems with those input features have successfully adapted some image data augmentation techniques (e.g. random cropping, scaling, etc.) \cite{hendrycks_2019:BenchmarkingNeuralNetwork}, the best performing methods use augmentation techniques designed for audio content, such as mixup with equalization EMDA \cite{takahashi2017aenet}  or without equalization \cite{zhang2017mixup}, spec augment \cite{park_2019:SpecAugmentSimpleData}, impulse response simulations \cite{Shimada_SONY_task3_report}, pitch and/or time shifting and stretching \cite{schluter2015exploring, salamon_2017:DeepConvolutionalNeural}, filtering, dynamic range compression \cite{salamon_2017:DeepConvolutionalNeural}, or spatial soundfield rotations \cite{mazzon_2019:FirstOrderAmbisonics}. 

Except for soundfield rotations and impulse response simulations, most of the aforementioned augmentation techniques were designed for monophonic or single channel audio, and only a few exploit the spatial characteristics of the input signals. To address this issue, in this paper we propose Spatial Mixup, which uses a general parametric spatial audio effect as a data augmentation technique by applying a directional loudness modification to the audio data. This modification effectively transforms the spatial characteristics of the recorded soundfield, enhancing sound arriving from some directions while suppressing others. However, unlike a beamformer, the overall transformation is gentle, such that the overall content (class) and DOA of the all recorded events is preserved. Figure \ref{fig:approach} shows a visualization of this concept compared with an omnidirectional 2D signal, and a traditional beamformer. The left column shows the omnidirectional responses, and the right column shows the the activity-coupled Cartesian DOA~(ACCDOA) vectors \cite{shimada_2021:ACCDOAActivityCoupledCartesian}, which assign an event activity to the length of corresponding Cartesian DOA vectors.

\section{Related work}

Soundfield rotations in ambisonics domain such as swapping, arbitrary rotation, rotation over a single axis were first proposed by \cite{mazzon_2019:FirstOrderAmbisonics}. These generate new DOA labels, that might not exist in the dataset. However, the overall soundfield stays constant, where the acoustic environment (i.e the room) is rotated as well, so the relationship between direct sound, early reflections, and reverberation remains unchanged. The relative positions between events is also preserved. Nonetheless, this has proven successful, especially for the localization subtask, at least when there are few overlapping sources. 

Expanding the concept of soundfield manipulations, \cite{wang_2021:FourStageDataAugmentation} proposed audio channel swapping (ACS) and multi-channel simulation (MCS), where ACS applies a similar soundfield rotation as the channel swapping in \cite{mazzon_2019:FirstOrderAmbisonics}, to both MIC and FOA signals. MCS aims to simulate new spatial information for specific events. To do this, a preprocessing step analyzes the recording to distinguish between noise and possible sources, then a beamformer extracts the direct sound while the spatial characteristics are extracted computing the spatial covariance matrix. This is comparable to a parametric decomposition into direct and diffuse components \cite{mccormack_:ParametricSpatialAudio, politis_2012:ParametricSpatialAudio}. Augmentation occurs by adding random perturbations to the spatial components, preserving the content, simulating new acoustical environments.


Finally, techniques that combine the content of multiple input signals have shown effectiveness too. First, mixup (referred in this paper as regular mixup) is a linear combination of an original signal with some other  signal. In the case of time domain audio signals, this is equivalent to mixing two sound tracks together, which translates into two sound events occurring at the same time. In addition, the labels corresponding to the signals can be combined too if available. The regular mixup \cite{zhang2017mixup} can be expressed as
\begin{equation}
\begin{gathered}
\boldsymbol{\hat{x}} =  \lambda \boldsymbol{x}  + (1 - \lambda) \boldsymbol{y}  ,\\
\end{gathered}
\end{equation}
where $\boldsymbol{x} $ is the single channel input audio signal, $\boldsymbol{y} $ is the interfering signal,  $\boldsymbol{\hat{x}}$ is the augmented signal, and $\lambda \sim Beta(\alpha, \beta) $ is a hyperparameter that controls the strength of the mixup. An alternative mixup is EMDA \cite{takahashi2017aenet}, applies random equalization to each signal, reducing the overlap in frequency domain, while preserving time mixing.

\section{Proposed method: Spatial Mixup}

For this paper, we assume that the input signals represent a soundfield encoded in ambisoncis format \cite{pomberger2009ambisonics}. This means that a full 3D sound scene has been captured by a microphone array and properly encoded into an orthonormal basis of spherical harmonics representing the full soundfield. 

The main idea of spatial mixup is to slightly modify the spatial characteristics of a recorded spatial audio signal to increase the robustness of neural networks models to these transformations. While regular mixup combines the content of two different sound signals, spatial mixup can be understood as applying the mixup operation to a spatially transformed version of the same signal. This operation is now defined as
\begin{equation}
\begin{gathered}
\Xhat =  \lambda \boldsymbol{X}  + (1 - \lambda) \mathcal{T} \boldsymbol{X} ,  \\
\end{gathered}
\end{equation}
where $\Xhat \in \mathbb{R}^{{n}_{\rm{out}}}$ is the augmented audio signal, ${n}_{\rm{out}} = (N_\text{out}+1)^2$ for the output order $N_\text{out}$, $\boldsymbol{X} \in \mathbb{R}^{{n}_{\rm{in}}}$ is the multi-channel spatial audio input signal, ${n}_{\rm{in}} = (N_\text{in}+1)^2$ for the input order $N_\text{in}$, and $\mathcal{T} \in \mathbb{R}^{n_{\text{out}} \times n_{\text{in}}} $ is a transformation matrix that performs linear combinations of the channels in $\boldsymbol{X}$ via a matrix multiplication. The transformation matrix $\Tmat$ is further decomposed as 
\begin{equation} \label{eq:Tmat}
\begin{gathered}
\Tmat = \Ygrid \boldsymbol{G} \boldsymbol{W} ,\\
\end{gathered}
\end{equation}
where $\Ygrid \in \mathbb{R}^{n_{\text{out}} \times n_{\text{grid}}}$ is a matrix of real spherical harmonics \cite{rafaely_2015:FundamentalsSphericalArray} of order $N_\text{out}$ computed at the azimuth and elevation of a discrete set of points sampled from a unitary sphere; $\boldsymbol{G} = \text{diag}[g_i] \in \mathbb{R}^{{n}_{\text{grid}} \times {n}_{\text{grid}}}$ is a diagonal matrix composed of gains for each point $i$ in the grid defined for $\Ygrid$ ; and $\boldsymbol{W} \in \mathbb{R}^{n_{\text{grid}} \times n_{\text{in}}}$ is a beamforming matrix that couples the number of input channels to the grid directions. This effectively spatially decomposes the input soundfield into a discrete sampling. Although $\boldsymbol{W}$ can be set to any beamforming matrix, in practice it is sufficient to set $\boldsymbol{W} = \frac{1}{(N_{in}+1)^2} \Ygrid^T$, which corresponds to a hypercardiod beamforming to the grid points.


The setup presented for $\Tmat$ can be applied to many spatial transformations depending on the values of  $\boldsymbol{G}$ , including warping, compression, and acoustic zooms \cite{zotter_2019:AmbisonicsPractical3D, kronlachner_:SpatialTransformationsEnhancement, politis_2012:ParametricSpatialAudio}. An additional rotation matrix $\boldsymbol{R}$ can be added to Equation \eqref{eq:Tmat} if needed. In this paper we focus the analysis on a modification known as directional loudness.

\begin{figure*}[htb]
    \hspace{2cm}
    \begin{minipage}[b]{0.33\linewidth}
      \centering
      \includegraphics[trim={0 2.5cm 0 1.8cm},clip, width=1\textwidth]{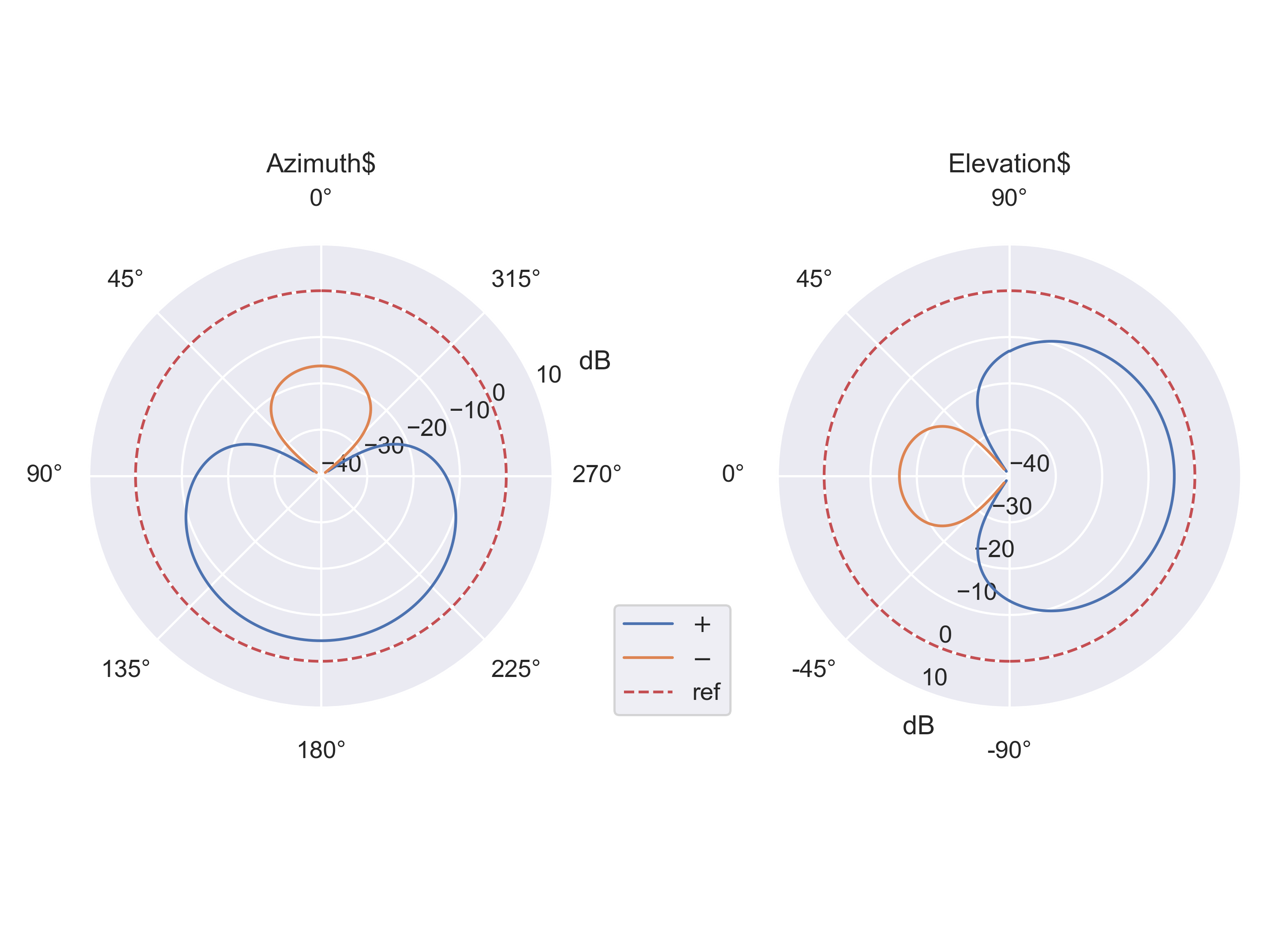}
    \end{minipage}
    \hspace{2cm}
    \begin{minipage}[b]{0.35\linewidth}
      \centering
      \includegraphics[trim={0.5cm 0.5cm 0 1.0cm},clip, width=1\textwidth]{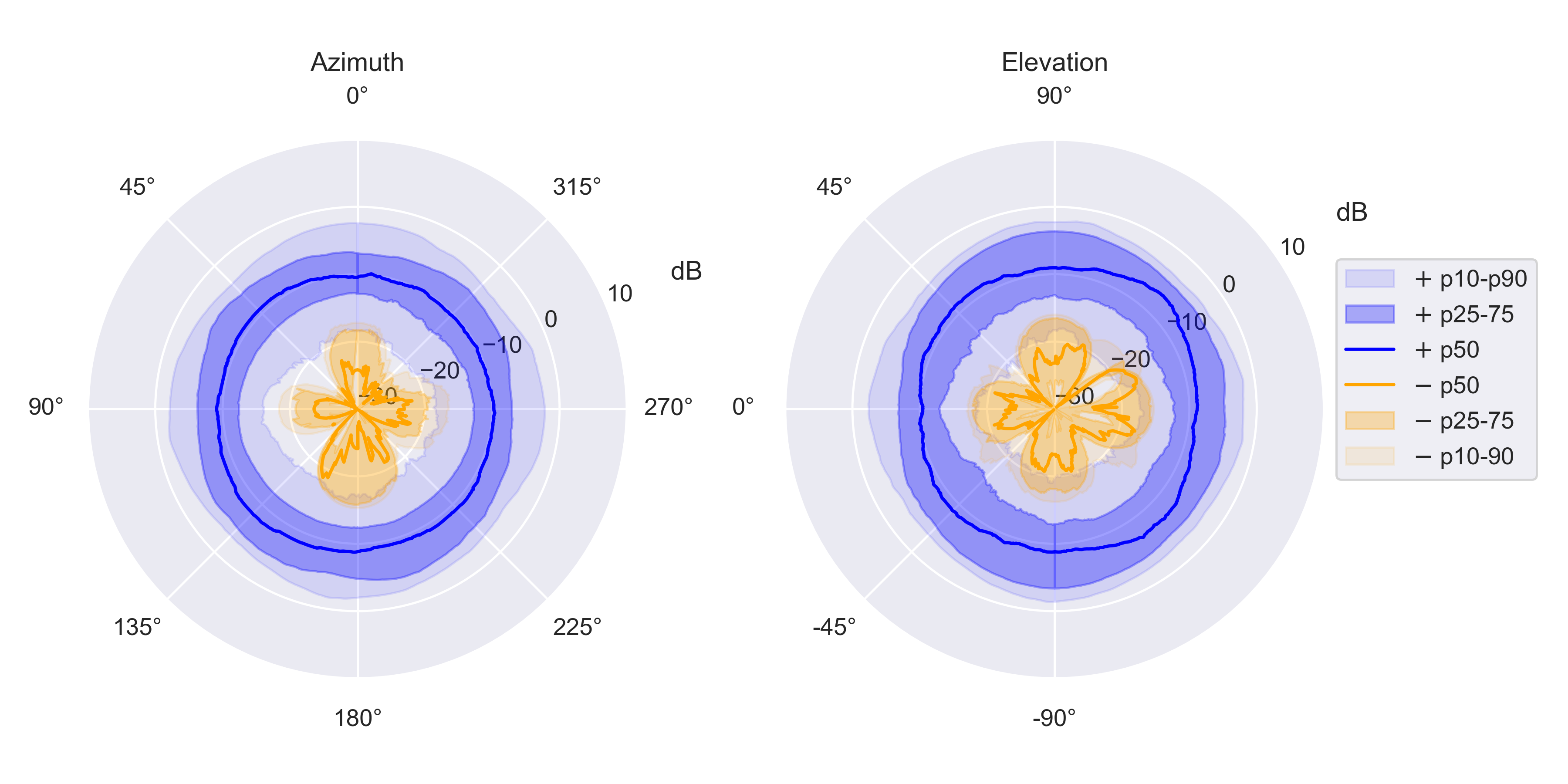}
    \end{minipage}
    
    \hspace{2cm}
    \begin{minipage}[b]{0.33\linewidth}
      \centering
      \includegraphics[trim={0 2.5cm 0 1.8cm},clip, width=1\textwidth]{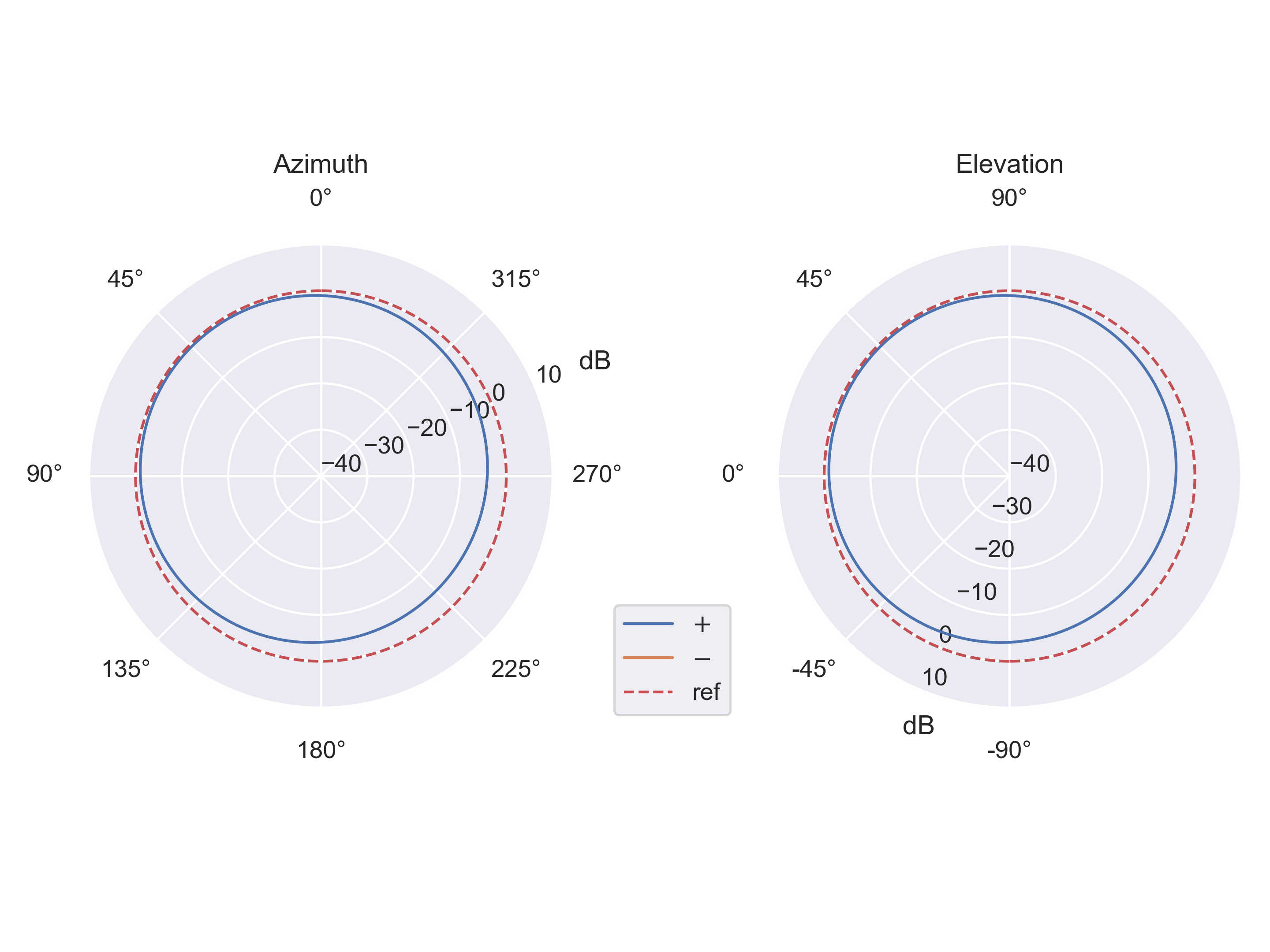}
    \end{minipage}
        \hspace{2cm}
    \begin{minipage}[b]{0.35\linewidth}
      \centering
      \includegraphics[trim={0.5cm 0.5cm 0 1.0cm},clip, width=1\textwidth]{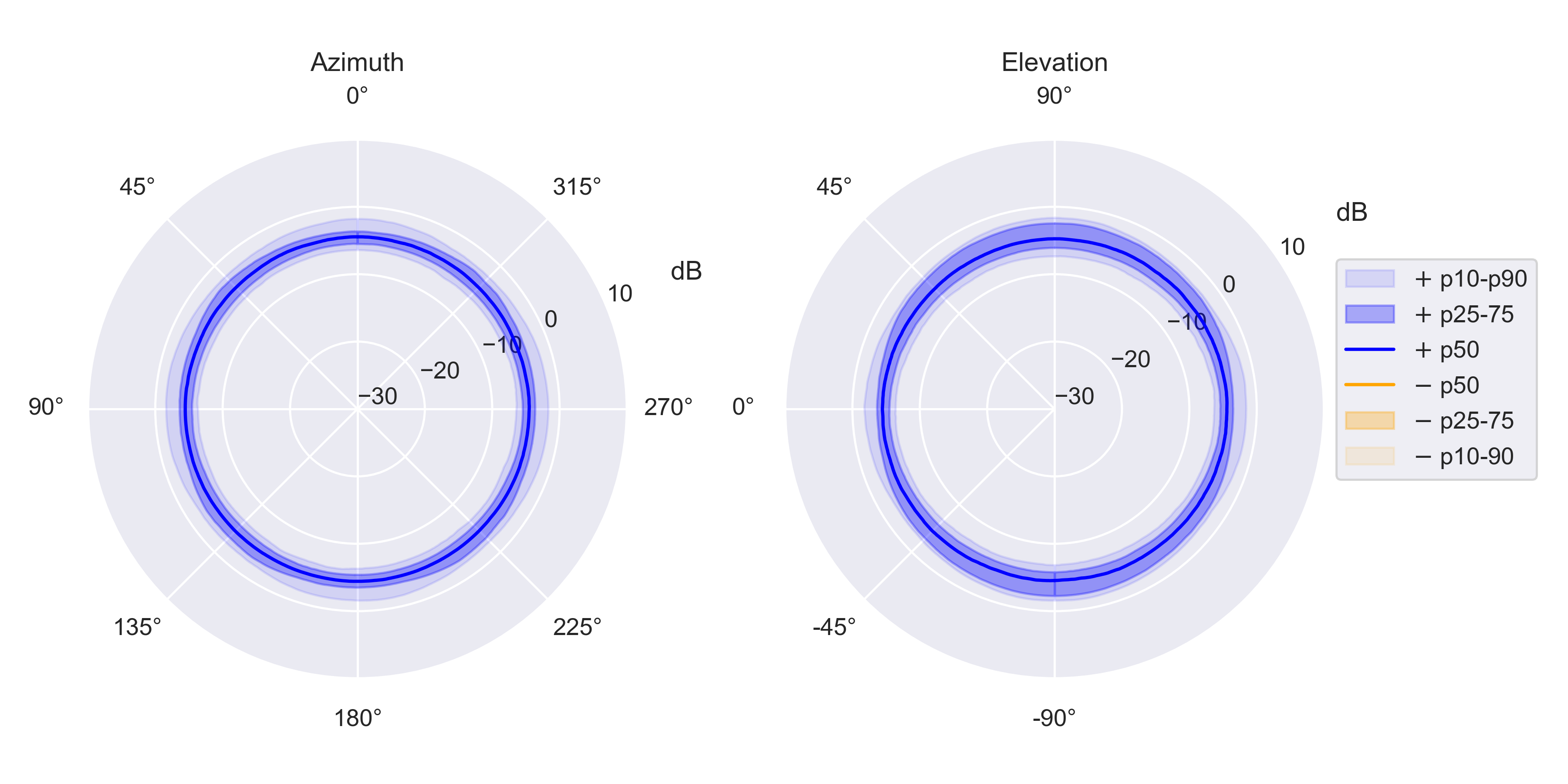}
    \end{minipage}
    \captionsetup{width=0.95\linewidth}
    \caption{Comparison of cross sections of the polar pattern responses for the omnidirectional channel of hard (top row) and soft (bottom row) spherical caps of first order. (Left column) A typical example directional loudness transform of each spherical cap. (Right) Distribution of the same responses for a sample of 500 patterns with randomized parameters for the spherical caps.}
    \label{fig:distribution}
    \vspace{-3mm}
\end{figure*}

\subsection{Directional Loudness}

A directional loudness modification corresponds to a spatial filter using some function. The spherical cap function divides the area of a unitary sphere into two sections, where the inner section is denoted as a spherical cap. This cap is parametrized by a central angle $\Omega_c = (\theta, \phi)$ of azimuth and elevation, and a width $\gamma_c$.  The spherical cap function \cite{kronlachner_:SpatialTransformationsEnhancement, zotter_2019:AmbisonicsPractical3D} is defined as the vector of gains
\begin{equation}
\begin{gathered}
g_i =  g_1 U(\Omega_c^T \Omega_i - \cos\frac{\gamma_c}{2}) + g_2 U( \cos\frac{\gamma_c}{2} - \Omega_c^T \Omega_i) ,\\
\end{gathered}
\end{equation}
for all $i \in  n_{\text{grid}}$ grid directions in $\Gmat$, $U$ is the unit step function, $\gamma_c$ is the width of the cap in radians, and $g_1$, $g_2$ are the gains for the region inside and outside the cap respectively.

Moreover, the aforementioned transformation matrix $\Tmat$ can also be applied to the SELD labels, especially if they are expressed as ACCDOA vectors. In this case, given that the ACCDOA labels are unitary activity vectors pointing to the DOA of a sound events, the application of spatial mixup applies $\Tmat$ to the labels is expressed as
\begin{equation} \label{eq:label}
\begin{gathered}
\boldsymbol{\hat{Z}} =  \lambda \boldsymbol{Z} + (1 - \lambda) \Tmat  \boldsymbol{Y}_0^{0} ,\\
\end{gathered}
\end{equation}
where $\boldsymbol{\hat{Z}}$ are the augmented labels, $\boldsymbol{Z}$ are the true labels expressed as ACCDOA vectors, and  $\boldsymbol{Y}_0^{0}$ are the spherical harmonics of first order and degree, computed for the direction of all active labels.

Generally, any function can be applied to $\Gmat$, but we experimentally found that the success of the directional loudness as augmentation method relies heavily on a transformation that is gentle and not too extreme. For this reason, we select two hyperparameters sets defined as \textit{soft} and \textit{hard} spherical caps, detailed in Table \ref{tab:hyper}. Figure \ref{fig:distribution} compares these two types of spherical caps, for both a single example of a typical response as well as the coverage and  distribution of a random sample of them. Overall, hard spherical caps cover a wider range and some patterns include negative phase regions, while soft spherical caps are more even, with smaller variations.

\setlength\tabcolsep{5.5pt} 
\begin{table}
    \footnotesize
    \captionsetup{width=0.98\columnwidth}
    \centering
    \caption{\label{tab:hyper}Hyperparameters for the spherical caps.}
    \vspace{-3mm}
    \begin{tabularx}{\columnwidth}{l l l c}
    \specialrule{1.5pt}{2pt}{3pt} 
    $\boldsymbol{G}_{type}$ & Parameter	& Distribution & Values	\\
    \specialrule{0.5pt}{2pt}{2pt}
    \multirow{ 5}{*}{Sph-Cap (Soft)}  &    Cap-center-azi  &   uniform & $[0, \pi]$\\
     &    Cap-center-ele  &   uniform &$[-\pi, \pi] $ \\
     &    Cap-width  &   uniform &$[\pi / 4, \pi] $ \\
     &    $G_1$  &   exponential &$[0, -3] $ \\
     &    $G_2$  &   uniform &$[-3, -6] $\\
    \specialrule{0.25pt}{1pt}{1pt}
         \multirow{ 5}{*}{Sph-Cap (Hard)}  &    Cap-center-azi  &  uniform &$[0, \pi]$\\
     &    Cap-center-ele  &   uniform &$[-\pi, \pi]$ \\
     &    Cap-width  &   uniform &$[\pi / 4, \pi/ 2]$ \\
     &    $G_1$  &   exponential &$[0, -6]$ \\
     &    $G_2$  &   uniform &$[-6, -20]$ \\
    \specialrule{1.5pt}{2pt}{10pt}
    \end{tabularx}
    \vspace{-3mm}. 
\end{table}


\section{Experiments}
\subsection{Experimental Setup}

To evaluate the performance of our method, we used the TAU-NIGENS Spatial Sound
Events 2021, introduced in the DCASE 2021 Task3 challenge \cite{politis_2021:DatasetDynamicReverberant}. This dataset includes 600 one minute long clips of spatial audio recordings, presented in two formats: the raw microphone array signals (MIC) and first order ambisonics (FOA), sampled at 24 kHz. The dataset is split into training, validation, and test subsets, consisting of 400, 100 and 100 minutes respectively. Each clip contains a dynamic acoustic scene, where specific sound events, are mixed together in a simulated acoustical environment. In total, there are 12 sound event classes, with examples such as footsteps or a dog barking. Each scene is generated with a collection of up to 3 concurrent sound events, that can be either spatially stationary, or follow a trajectory inside the sound scene. In addition, interference noise along with background noise are also present, where the former are localized sound events that do not belong to any of the classes, and the latter are continuous multi-channel recordings of ambient noise naturally present in the acoustical environments where the impulses were collected.

The evaluation tasks hence consists of the classification of sound events as well as the estimation of DOAs for full clips. For this purpose we used the same classification (\metricA, \metricB) and localization (\metricC, \metricD) as explained in the official DCASE challenge \cite{politis2020overview}. 
We also adopted an aggregated SELD error (\metricE), as
\begin{align}
    \rm{\mathcal{E}_{SELD}} = \frac{\rm{{ER}_{LD}} + ( 1 - \rm{{F}_{LD}} ) + \frac{\rm{{LE}_{CD}}}{\pi} + ( 1 - \rm{{LR}_{CD}} )}{4}.
    \label{eq:seld_error}
\end{align}

The goal of the experiments was to compare the impact of the augmentation methods fairly, and not necessarily to get the best possible performance in the task. Consequently, the training setup was the same for all. All experiments were trained for 100,000 iterations of batch size 32, with validation every 10,000 steps, using the official subset split. We minimize the MSE loss, using Adam optimizer and a learning rate scheduler with a warmup stage starting at learning rate 1e-4, reaching 1e-3 after 5 validation steps, followed by a reduce on plateau scheduler (monitoring the validation SELD error) with patience of 3 validation steps and a decay rate of 0.9. For each experiment, we report the test subset results, from the model of the best validation step of multiple runs.

\subsection{Features and models}

The experiments are conducted using two different systems, a low complexity model with standard architecture, and another with a more sophisticated model with a large amount of parameters. The systems are:
\begin{enumerate}
    \item \textbf{Basic system} - We use the CRNN10 model proposed by \cite{caoCRNN}, which consists of 2d convolutional layers with batch normalization and increasing number of channels. The inputs are linear amplitude STFT and interchannel differences using only the FOA input signals, for a total of 7 input channels. The spectrograms are computed using frame size of 512, hop size of 240, and total input length for the network is 1.27 seconds. 
    
    \item \textbf{Sophisticated system} - We use the RD3Net \cite{shimada_2021:ACCDOAActivityCoupledCartesian}, which consists of a series of densely connected blocks of 2d, dilated convolutions. Each block is followed by a down sampling module and finally, a gated recurrent unit (GRU), fully-connect (FC) layer, and up sampling operation as outputs. The input features are the same as in the basic system.
\end{enumerate}

\subsection{Results}

\subsubsection{Effects of Directional Gain $ \Gmat $ }

Table \ref{tab:one} shows the performance of the proposed augmentation method with different types of $ \Gmat $ compared with a baseline without any data augmentation for the basic model. In these experiments, the spatial mixup was applied only to the input signals $\boldsymbol{X}$, the output order was 1, and the  $\Ygrid $ was computed for a t-design \cite{graf2011computation} of degree 3 (giving 6 total directions).  For most metrics, the performance is better for all $ \Gmat $ types except random (uniformly random diagonal matrix), which is significantly worse. Surprisingly, the identity matrix and the soft spherical caps show similar results, while the hard spherical caps are not as good. The former can be explained in part because the full transformation matrix is not truly orthogonal, due to the limited spatial resolution in the grid that generates some small non-zero values. The latter is most likely because the hard spherical caps can sometimes generate extreme patterns that resemble a beamformer, rather than a gentle manipulation of the soundfield. These patterns are possibly too extreme for the model to learn invariance.

\setlength\tabcolsep{3.5pt} 
\begin{table}
    \footnotesize
    \captionsetup{width=0.98\columnwidth}
    \centering
    \caption{\label{tab:one}Performance of Spatial Mixup with different directional loudness $\Gmat$ matrix types in the \textit{basic} system.}
    \vspace{-3mm}
    \begin{tabularx}{\columnwidth}{l c c c c c }
    \specialrule{1.5pt}{2pt}{3pt}
    System & \metricA   & \metricB  & \metricC  & \metricD  &   \metricE \\
    \specialrule{0.5pt}{2pt}{2pt}
    Baseline   & 0.689 & 40.5 & 20.7 & 44.4 & 0.489 \\
    Random       & 0.776    &   24.2    &   26.9    & 32.5  &   0.590 \\
    Identity	 & 0.668    & \textbf{42.2}  & 19.5  & 42.9  & 0.481 \\
    Sph-Cap (Hard) &	0.693   &   39.1    & 22.1  & \textbf{45.6}  & 0.492 \\
    Sph-Cap (Soft) &	 \textbf{0.664}  & 42.1  &    \textbf{19.4}   &    43.2   &    \textbf{0.480} \\
    \specialrule{1.5pt}{2pt}{10pt}
    \end{tabularx}
    \vspace{-3mm}
\end{table}

The same comparison of different $\Gmat$ types was explored for the sophisticated model in Table \ref{tab:two}. Here, the spatial mixup setup was the same as the previous experiments, except that the  $\Ygrid $ was computed with a larger t-design of degree 7, for 24 directions, giving better spatial resolution. The results show a similar trend as Table  \ref{tab:two}, where the main difference is that the $ \Gmat $ type identity was not quite as good as the soft spherical caps. This suggests that the small non zero values are not as strong here, due to the larger number of directions in the grid. In addition, the higher capacity of the model is able to accommodate for the soft spherical caps, learning better invariance. In summary, a smooth $\Gmat$ function, with a sufficiently large grid works best.

\setlength\tabcolsep{3.5pt} 
\begin{table}
    \footnotesize
    \captionsetup{width=0.98\columnwidth}
    \centering
    \caption{\label{tab:two}Performance of Spatial Mixup with different directional loudness  $\Gmat$ matrix types in the larger \textit{sophisticated} system. }
    \vspace{-3mm}
    \begin{tabularx}{\columnwidth}{l c c c c c }
    \specialrule{1.5pt}{2pt}{3pt}
    System & \metricA   & \metricB  & \metricC  & \metricD  &   \metricE \\
    \specialrule{0.5pt}{2pt}{2pt}
    Baseline	 & 0.678   &	43.5   &	21.8  &	53.5   &	0.458 \\
    Random       & 0.744	&   27.0  &	29.3   &	41.8   &	0.555 \\
    Identity	 & 0.643	&   46.9   &	22.1   &	\textbf{56.0}   &	0.434 \\
    Sph-Cap (Hard) 	& 0.660    &	44.7   &	22.1   &	55.7   &	0.445 \\
    Sph-Cap (Soft) 	& \textbf{0.615}    & \textbf{48.9}    &	\textbf{18.7}   &	54.2   &	\textbf{0.422} \\
    \specialrule{1.5pt}{2pt}{10pt}
    \end{tabularx}
    \vspace{-3mm}
\end{table}

\subsubsection{Other effects}

The general method allows to utilize a different order for the outputs than the input signals. However, even if a higher output order is selected, it is not possible to generate spatial information that is not already there. That said, for data augmentation purposes, having a higher order might enable a different architecture for neural network models, (e.g. with more input channels), as well as more nuanced directional loudness modifications. In addition, the labels can also be augmented as described in Equation \eqref{eq:label}. Cursory experimental results of both effects showed very small differences to results already presented, so these are not explicitly included in this paper. Nevertheless, it is possible that other tasks, in particular tasks with high order ambisonics data might see more significant performance gains when using spatial mixup.

\subsubsection{Comparison to other data augmentation }

\setlength\tabcolsep{3.5pt} 
\begin{table}
    \footnotesize
    \captionsetup{width=0.98\columnwidth}
    \centering
    \caption{\label{tab:four}Performance of common data augmentation method compared with the Spatial Mixup using the \textit{basic} system.}
    \vspace{-3mm}
    \begin{tabularx}{\columnwidth}{l c c c c c }
    \specialrule{1.5pt}{2pt}{3pt}
    System & \metricA   & \metricB  & \metricC  & \metricD  &   \metricE \\
    \specialrule{0.5pt}{2pt}{2pt}
    Baseline (B1)   & 0.689 & 40.5 & 20.7 & 44.4 & 0.489 \\
    B1+Mixing	    & 0.649 & 45.7 & 20.4 & 51.9 & 0.447 \\
    B1+Rotation	    & 0.633 & \textbf{46.5} & 20.4 & 51.1 & 0.442 \\
    B1+SpecAugment  & 0.702 & 37.6 & 23.4 & 45.2 & 0.501 \\
    B1+EQ	        & 0.675 & 42.5 & 20.9 & 44.6 & 0.480 \\
    B1+All          & 0.652 & 46.2 & 22.4 & \textbf{57.3} & \textbf{0.435} \\
    \specialrule{0.5pt}{1pt}{1pt}
    B1+Sph-cap (Soft)	    & 0.662 & 42.8 & \textbf{20.1} & 45.7 & 0.472 \\
    B1+All+Sph-cap (Soft)  & \textbf{0.628} & 46.3 & \textbf{20.1} & 50.8 & 0.442 \\
    \specialrule{1.5pt}{2pt}{10pt}
    \end{tabularx}
\end{table}

Table \ref{tab:four} compares spatial mixup to other common data augmentation methods including mixing \cite{shimada_2021:ACCDOAActivityCoupledCartesian}, spec augment \cite{park_2019:SpecAugmentSimpleData}, FOA soundfield rotations \cite{mazzon_2019:FirstOrderAmbisonics}, random equalization \cite{takahashi2017aenet} and a combination of all four.  These results show that from the common augmentations, FOA rotations and mixing are the best performers, achieving a faster convergence and better metrics in general. While random EQ increases the performance slightly, it is surprising that spec augment is markedly worse than the baseline. More importantly, spatial mixup with soft spherical caps (using a grid with t-design of degree 7) shows considerable performance gains over the baseline, better than EQ but slightly less than mixing. Lastly, when comparing the combinations with and without spatial mixup, it seems that adding spherical caps reduces errors, but also reduces recall for events. However, larger systems might exploit this better.

A possible explanation for these results is that the DCASE 2021 Task3 benefits the most from methods that improve equivariance rather than invariance, given that both FOA rotations and mixing increase the coverage of the labels. In contrast, spatial mixup modifies the relative levels of certain directions, increasing sound level diversity for all events. Nonetheless, spatial mixup shows good results.


\section{Conclusions}


In this paper, we proposed a data augmentation method for sound event localization and detection (SELD) tasks, based on the application of spatial audio parametric effects, in a process we call Spatial Mixup. This enables modifications to the spatial characteristics of audio signals encoded in ambisonics format, by applying a transformation matrix to the time domain input signals. This matrix is obtained by spatially sampling the original soundfield, and appliying a directional loudness gain modification. The method was evaluated in the DCASE2021 Task3 dataset, which includes complex sound scenes with overlapping and non-stationary sources, as well as interference and background noise. The method proved effective when using gentle modifications known as soft spherical caps. It improves all metrics when compared to a non-augmented baseline, and shows similar advantages compared to well known augmentation methods. Future research could explore the application of the method for other machine learning tasks with spatial audio, as well as analyze further transforms such as audio warping or acoustic zoom.

\vfill\pagebreak

\clearpage
\newpage

\bibliographystyle{IEEEbib}
\bibliography{myRefs}

\end{document}